# Hourly Forecasting of Emergency Department Arrivals– Time Series Analysis

Avishek Choudhury
School of Systems and Enterprises
Systems Engineering – Sociotechnical
Stevens Institute of Technology
Hoboken, NJ

## ABSTRACT

*Background:* The stochastic behavior of patient arrival at an emergency department (ED) complicates the management of an ED. More than 50% of a hospital's ED capacity tends to operate beyond its normal capacity and eventually fails to deliver high-quality care. To address the concern of stochastics ED arrivals, many researches has been done using yearly, monthly and weekly timeseries forecasting. *Aim:* Our research team believes that hourly time-series forecasting of the load can improve ED management by predicting the arrivals of future patients, and thus, can support strategic decisions in terms of quality enhancement. *Methods:* Our research does not involve any human subject, only ED admission data from January 2014 to August 2017 retrieved from the UnityPoint Health database. Autoregressive integrated moving average (ARIMA), Holt–Winters, TBATS, and neural network methods were implemented to forecast hourly ED patient arrival. *Findings:* ARIMA (3,0,0) (2,1,0) was selected as the best fit model with minimum Akaike information criterion and Schwartz Bayesian criterion. The model was stationary and qualified the Box–Ljung correlation test and the Jarque–Bera test for normality. The mean error (ME) and root mean square error (RMSE) were selected as performance measures. An ME of 1.001 and an RMSE of 1.55 were obtained. *Conclusions:* ARIMA can be used to provide hourly forecasts for ED arrivals and can be utilized as a decision support system in the healthcare industry. *Application:* This technique can be implemented in hospitals worldwide to predict ED patient arrival.

**Introduction**

Overcrowding in emergency departments (EDs) is acknowledged as a global concern. ED crowding can potentially hinder patient care by causing adjournments and medical errors during the treatment process. [1] [2] [3] Overcrowded EDs has resulted in deteriorating quality of care since 1987 [4]. The issue of overcrowded EDs was first addressed in a statewide conference on overcrowding held in New York City, encompassing the New York (NY) chapter of the American College of Emergency Physicians (ACEP), New York Emergency Medical Services (EMS), the NY State Department of Health, and state legislators. [4] In this conference, the problem was identified; however, no robust solution was introduced. [4] To address concerns related to overcrowded EDs, in this study, we defined measures for assessing overcrowding and conducted analyses to develop a robust solution.

Developing a scientific definition for overcrowded EDs is a complex problem and is a subject of debate. There is no definition that is based on specific wait times or quantitative permissible delays. American College of Emergency Physicians (ACEP) in 2006 defined ED overcrowding as "[the situation that] occurs when the identified need for emergency services exceeds available resources for patient care in the ED, hospital, or both." [5] This infers that overcrowding is a quantified relationship between service demand and capacity of delivering the desired service. Aspin in 2003 concluded that ED overcrowding is a measure of interdependence among ED input, ED throughput, and ED output. [6] Pines in 2007 attached the measure of quality of care with ED overcrowding. He proposed that "an ED is crowded when inadequate resources to meet patient demands lead to a reduction in the quality of care." [7] The absence of a scientific definition and measure of overcrowding makes it difficult to develop effective strategies to resolve ED problems.

ED overcrowding is triggered by various factors that include the patient population, physical capacity, practical capacity, pragmatic capacity, functional capacity, and fiscal capacity of the ED. Moreover, with the increase in population, ED demand in the United States has soared from 96.5 million to 115.3 million annual visits between 1995 and 2005. [8] [9] In contrast, during the same period, 381 EDs and 535 hospitals were closed. Not only did the gap between demand and the overall capacity of the ED increase, but superfluous ED [10] visits were also considered as significant factors responsible for this issue. Our study focuses on applying time series forecasting methods to solve ED overcrowding problems.

Time series forecasting methods has been used in various fields of healthcare, such as in forecasting daily outpatient visits, [11] inpatient admission, [12] maternal mortality, [13] emergency department visits, [14] disease management, [15] [16] healthcare waste generation, [17] and hospital census. [18] Additionally, time series forecasting methods are also used in various fields, including power and energy, [19] finance, [20] and traffic-flow. [21]

Existing studies have adopted several metrics such as National Emergency Department Overcrowding Score (NEDOCS), [22] and ambulance diversion [23] to measure and address ED overcrowding. However, these approaches reflect current or past ED conditions, and cannot predict upcoming ED visits and admits based on any statistical theory. There are limited studies that use different time series forecasting techniques to predict annual arrival, [24] monthly arrival, [14] [25] daily arrival, [26] and arrival at intervals of 4 to 12 h; [27] however, existing studies have shown complications in predicting hourly patient ED arrivals [28] and the application of predicting hourly patient arrivals is not well established. Therefore, this study introduces an hourly forecasting method to forecast ED arrivals with acceptable accuracy.

## Methodology

Our team retrospectively extracted ED arrival data from the EPIC and APPOLO databases used by UnityPoint Health in Des Moines city, Iowa. UnityPoint Health consists of three hospitals located in the state of Iowa. We collected hourly ED arrivals data from all the three hospitals from January 2014 through August 2017 for our time series analysis. The dataset was divided into training and actual outsample patient arrival datasets. The analysis was conducted on the training dataset from January 2014 through July 2017, and the forecasted values were matched against actual outsample patient arrival from January 2014 through August 2017.

During the process, the Kwiatkowski–Phillips–Schmidt–Shin (KPSS) test [29] was used to determine whether the time series was stationary around a mean or linear trend or is non-stationary owing to a unit root, and white noise was tested using the Box–Ljung test. [30] Moreover, we employed the Jarque–Bera test [31] and the Anderson–Darling test for normality, [32] and the augmented Dickey–Fuller test [29] to determine its stationarity. This was then followed by plotting of the autocorrelation function and partial autocorrelation function to determine ARIMA (p, d, q), where p is the order of autoregression (AR), d is the lagged difference between the current and previous values, and q denotes the order of the moving average (MA). The largest values of $p$ and $q$ were set as 24, and the maximum number of non-seasonal difference (d) was set as four [33]. Additionally, Holt–Winters, TBATS, and neural network (NN) methods were also implemented and compared with ARIMA.

The fitted model with minimum Akaike information criterion (AIC) and Schwartz Bayesian criterion (BIC) was selected as the optimal forecasting model. [14] The accuracy of the forecast was then measured based on the root mean square error and mean error.

**Results**

Figure 1 shows the original data decomposed into observed, trend, seasonal, and random data.

Figure 2 depicts the repetitive and moderate oscillation of seasonality with spikes at almost equal intervals. Figure 2a and Figure 2b shows the decomposed average 30-day plot and average 24-hour plot, respectively.

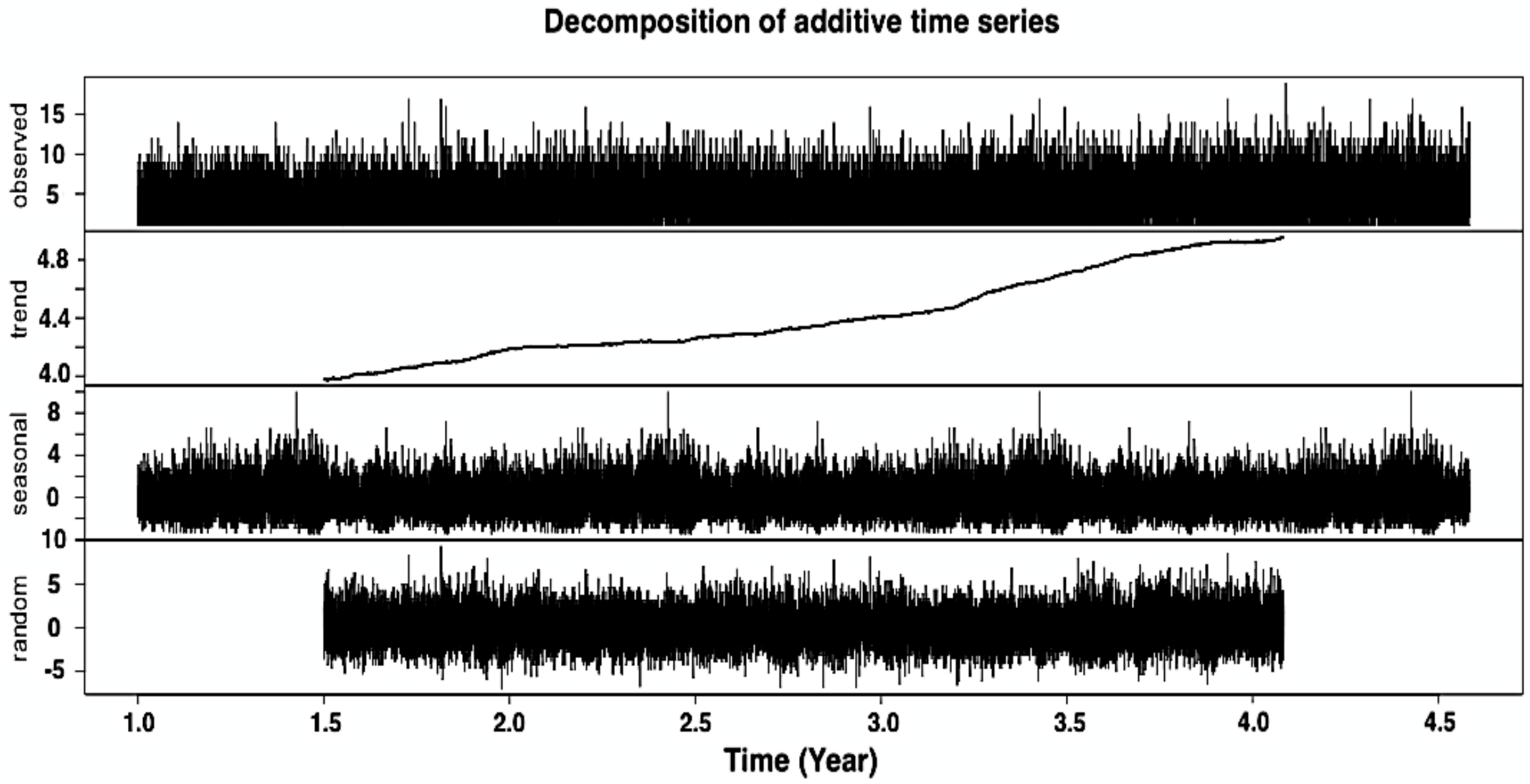

**Figure 1**: Decomposed time series plot of hourly ED visits from January 2014 through July 2017. The observation plot shows the increasing trend over the past four years and a repeating seasonal effect with one spike each year.

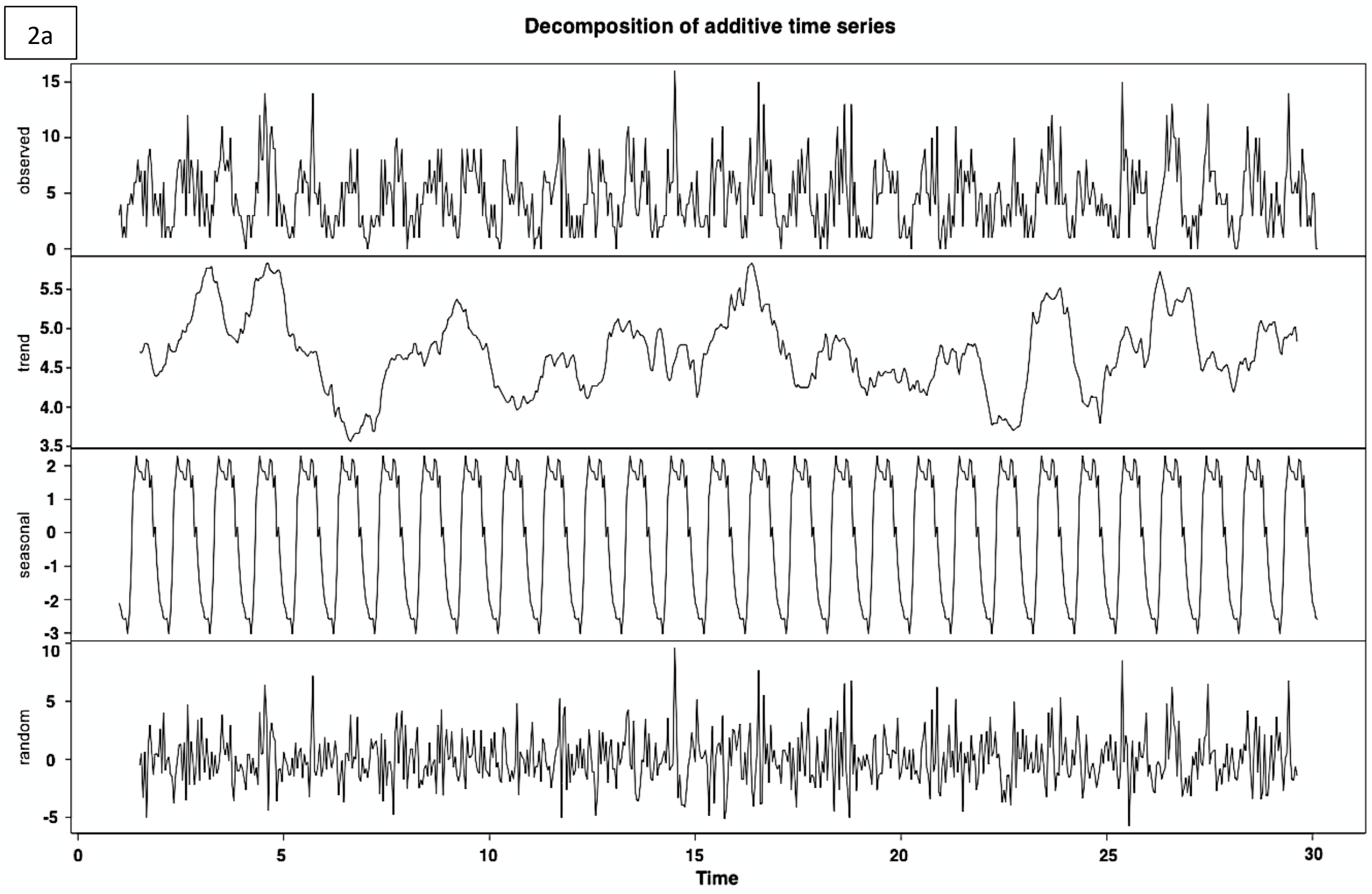

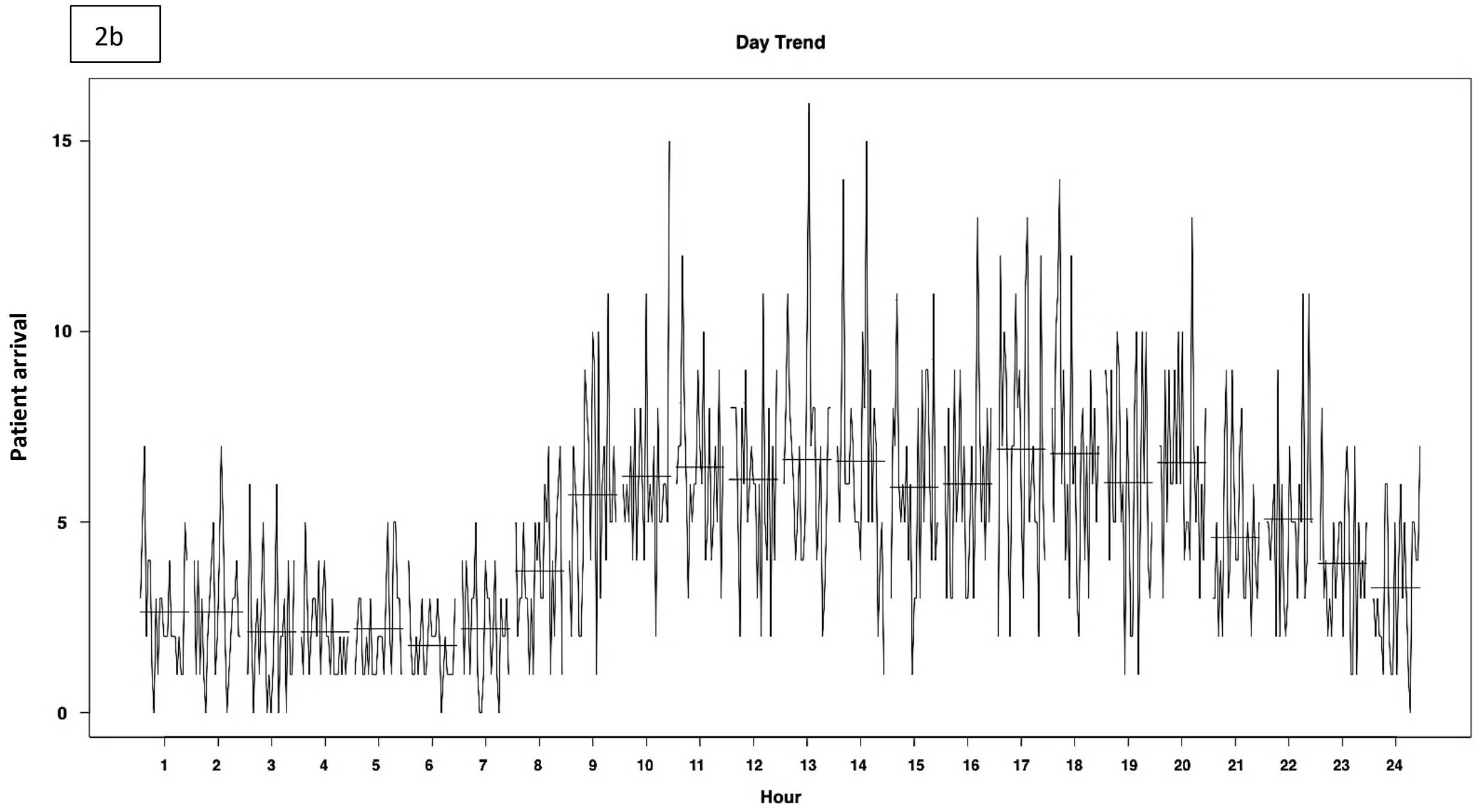

**Figure 2**: (a) Decomposed time series plot of daily ED visit over a 30-day period (January 2014–July 2017). The observational plot shows the fluctuating trend over 30 days and cyclic

seasonality. (b) Trend plot of daily ED visit over a 24-h period (average of January 2014–July 2017).

The best fit model ARIMA (3,0,0) (2,1,0)[24] was selected. Table 1 compares the ME and RMSE of TBATS, Holt–Winters, NN, and ARIMA. Table 2 describes the selected ARIMA model.

**Table 1**: Comparative analysis

| **PERFORMANCE MEASURE** | | | | | | | |
|---|---|---|---|---|---|---|---|
| **TBATS** | | **Holt-Winters** | | **Neural Net** | | **ARIMA** | |
| **ME** | **RMSE** | **ME** | **RMSE** | **ME** | **RMSE** | **ME** | **RMSE** |
| **1.75** | 2.28 | 1.19 | 27.86 | 1.4 | 3.26 | 1.001 | 1.55 |

Although TBATS and NN provides acceptable outcomes, ARIMA is selected as the best fit model. It exhibits the highest forecasting accuracy. NN is a strong time series forecasting method. However, wrong selection of network parameters might lead to overfitting [34] [35], thus producing good in-sample fit while hindering accurate forecasting.

**Table 2**: ARIMA model

| | **COEFFICIENTS** | | | | | **PERFORMANCE MEASURE** | |
|---|---|---|---|---|---|---|---|
| | **ar1** | **ar2** | **ar3** | **sar1** | **sar2** | **Mean Error** | **RMSE** |
| | 0.159 | 0.100 | 0.047 | -0.584 | -0.274 | 1.001 | 1.55 |
| **S.E** | 0.005 | 0.005 | 0.005 | 0.005 | 0.005 | | |

To ensure a good fit for ARIMA (3,0,0) (2,1,0),[24] we conducted the following tests as shown in the following table 3: **(a)** Jarque–Bera test, **(b)** Anderson–Darling normality test, **(c)** Box–Ljung test, and **(d)** Augmented Dickey–Fuller test.

**Table 3**: Model evaluation

| TEST | P- VALUE (<0.05) | INFERENCE |
|---|---|---|
| **Jarque-Bera test** | $2 \times 2e^{-12}$ | Rejects the null hypothesis of normality |
| **Anderson-Darling normality test** | $2 \times 2e^{-12}$ | Rejects the null hypothesis of normality |
| **Box–Ljung test** | 0.17 | No significant autocorrelation |
| **Augmented Dickey–Fuller test** | 0.01 | Rejects the null hypothesis of non-stationarity |

The Jarque−Bera test and Anderson−Darling normality test of residuals yielded a p-value of $2*2e^{-12}$ *(p-value <0.05)*. Thus, both the test rejects the null hypothesis of normality implying the residuals are not normally distributed.

The Box–Ljung test on residuals gave a p-value of 0.17 (p-value > 0.05), which implies that there is no significant autocorrelation. In other words, there is a lack of proof of independence. The augmented Dickey–Fuller yielded a p-value of 0.01 (p-value < 0.05); thus, the test rejects the null hypothesis of non-stationarity. (see figure 3a, 3b, and 3c below for ACF, PACF and Normality plot).

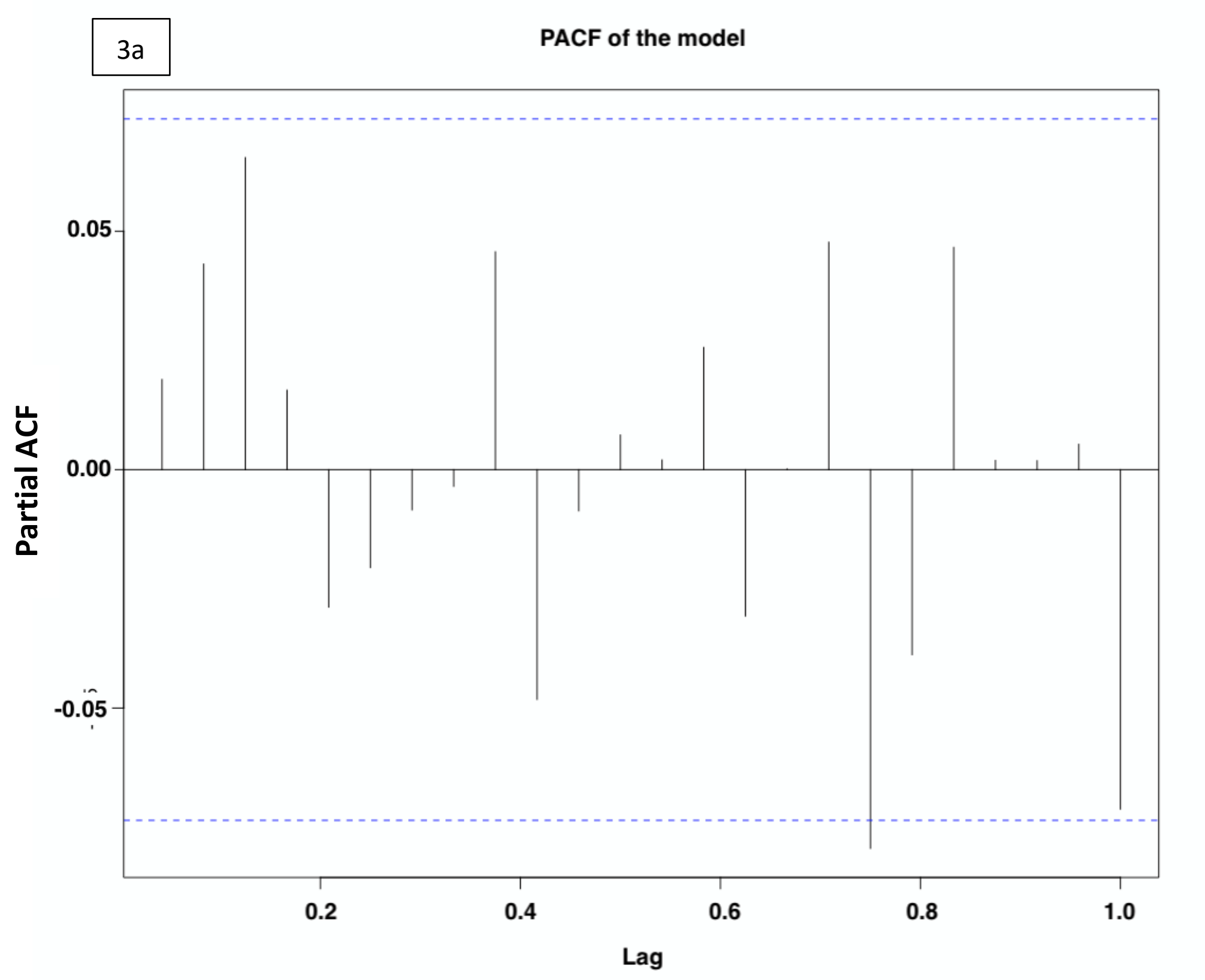

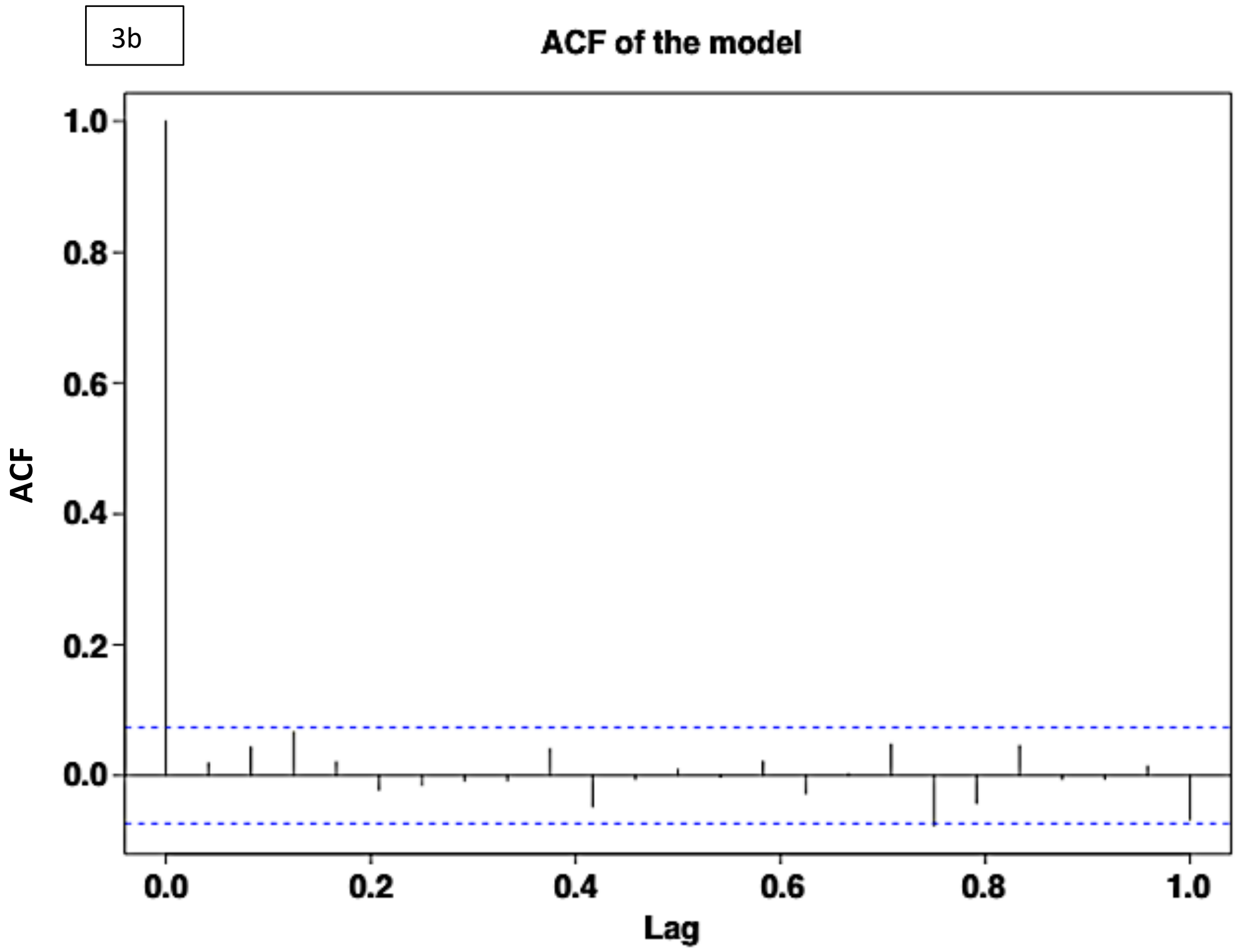

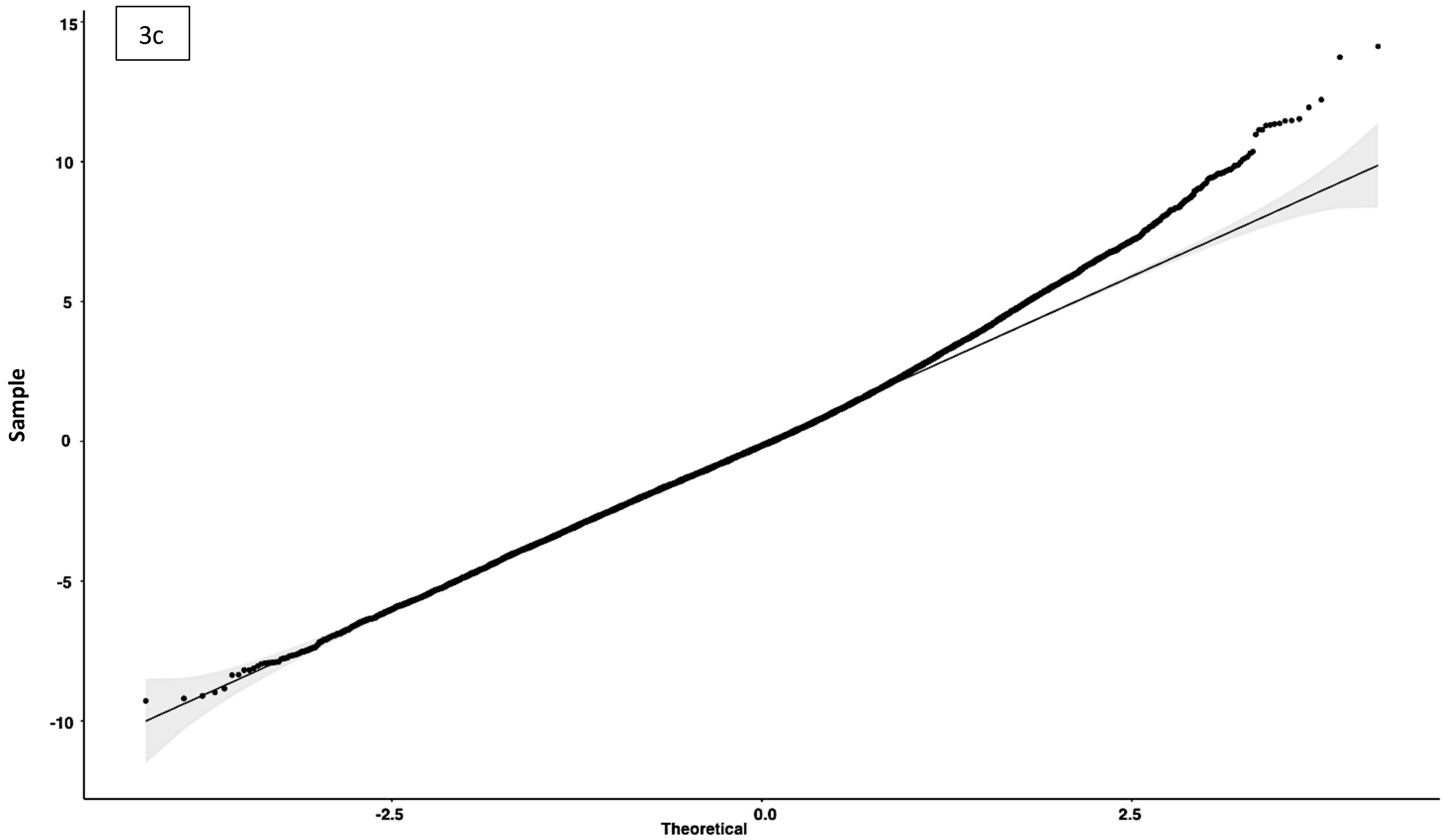

**Figure 3**: **(a)** shows the PACF plot with one point exceeding the lower limit at 0.8, **(b)** shows the ACF plot, and **(c)** shows the normality of the residuals (theoretical vs sample).

The best forecast accuracy was exhibited by ARIMA (3,0,0) (2,1,0).[24] The forecasted table is provided in supplementary file 2. The forecasted and actual values are compared and shown in Figures 4a and 4b.

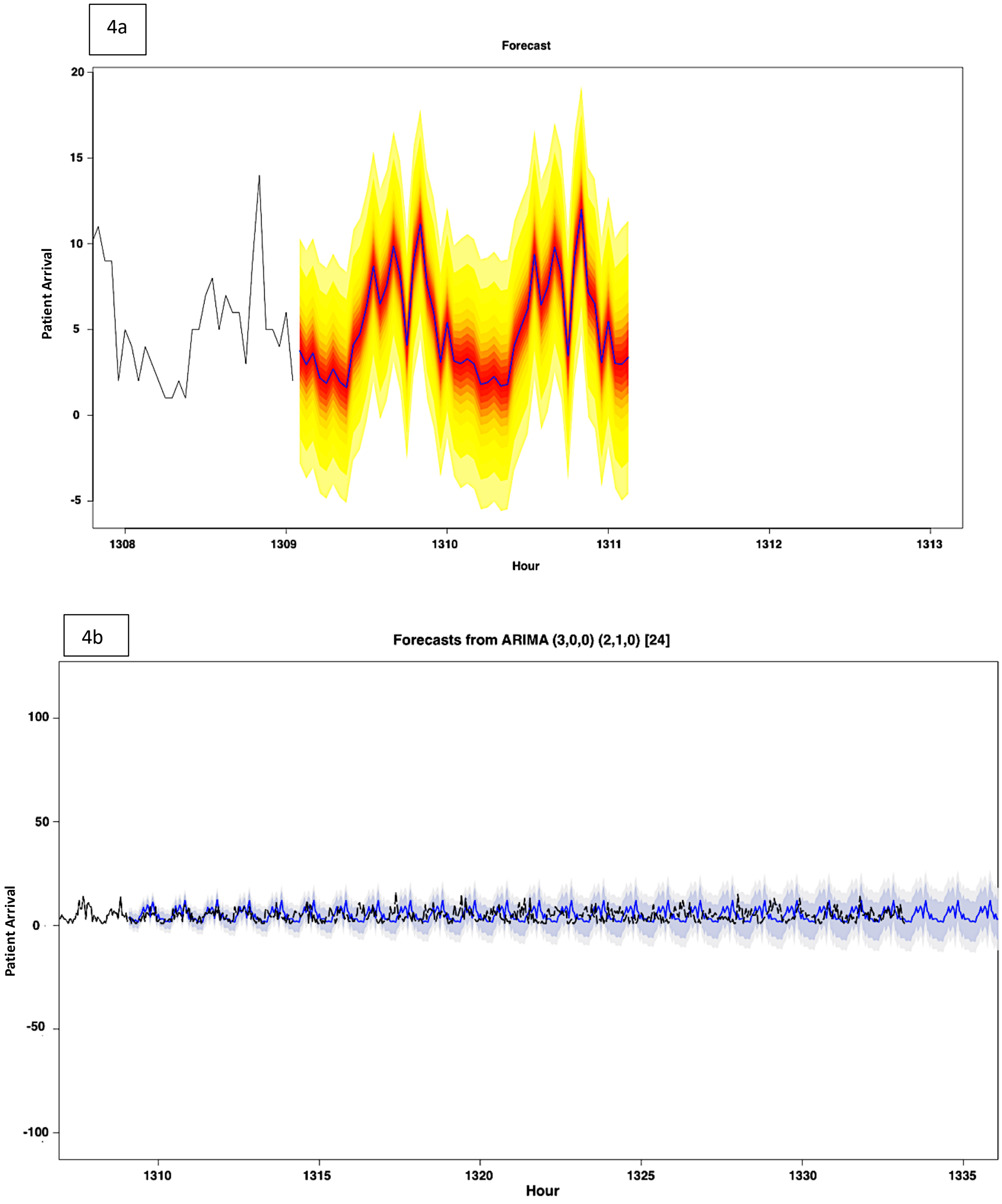

**Figure 4**: (a) Forecasted ED arrivals. The color shades in the figure show the different confidence intervals starting from 10% through 99%. (b) Actual ED arrivals (black line) versus Forecasted ED arrivals (blue line) at 80% and 95% upper and lower confidence intervals.

In Figure 4b, the solid blue line represents the forecasted ED arrival, whereas the dashed black line denotes the actual ED arrival. As shown in Figure 4b, all the forecasted values lie within the 80% and 95% upper and lower confidence intervals. By comparing the actual and forecasted values, we determined the mean error and root mean square error to be 1.001 and 1.55, respectively. Thus, the forecasting accuracy was high and acceptable. The forecasted ED arrivals were observed to increase with time, which holds good for managing the ED at UnityPoint Health.

**Discussion**

In the pursuit to moderate the unfavorable consequences of ED crowding, ED administrators must make improvements, including architectural enhancements and optimal staffing. To support these improvements, it is useful to forecast the input volume that the ED will need to process. The aptitude to foresee patient visits is imperative to expedite strategic decisions and quality improvements. To facilitate forecasting ED arrival, numerous methods have been applied. The time series analysis is simpler to use as time series models only use proof regarding the variable to be forecasted.

Time series analysis for forecasting ED arrivals have been pronounced in studies with acceptable results. [28] The ARIMA algorithm proposed in our study was able to accurately predict hourly ED visits, with an ME of 1.001 and an RMSE of 1.55. Based on error measures and visual interpretation of the results, we can infer that the prediction provides a useful assessment to predict hourly ED visits. Moreover, the use of the ARIMA algorithm in an open source platform makes time series analysis applicable for many hospitals and researchers.

Theoretically, ED administrators may use hourly forecasting of ED visits for adjusting staffing levels according to the projected clinical activity. With regard to regulating staffing levels, predictions of weekly or hourly ED visits would be preferable. [28] Existing literature have shown complications with regard to predicting daily and hourly patient ED visits. [28] There is limited research regarding the development of this staffing modification. [28] The number of patient arrivals affect the required level of staffing. [36] Morten Hertzum in 2017 implemented a time series forecasting model using ARIMA; however, this model did not test for residual normality, stationarity, and autocorrelation. [37] Our study not only tested the proposed model for normality, stationarity, and autocorrelation but also compared it with NN, Holt–Winters, and TABTS.

Future research regarding the application of hourly ED arrivals must therefore focus on incorporating unpredictable factors into the model and generalize its applicability to other fields of healthcare. As the number of visiting ED patients is often related to quality and performance, it could be useful to link this hourly prediction to quality assuring programs, scheduling ED staffs, and patient safety interventions.

**Strengths and limitations of this study**

•In this study, we developed a prediction model that has the highest prediction accuracy with respect to the prediction of hourly patient arrival and is the first time series forecasting study implemented in Iowa, USA.

•The study not only provides information on hourly patient arrivals, but also explains the yearly, monthly, and daily patterns of patient visits.

•This model not only forecasts ambulatory patient arrivals, but also considers arrivals in ambulances.

•The model developed in this study was implemented in forecasting patient arrivals in UnityPoint Health ED and Cath lab.

•This model does not consider any special circumstances (such as bad weather, interrupted transportation, and epidemics) that might influence ED arrivals.

## Conclusion

In this study, ARIMA (3,0,0) (2,1,0)[24] was found to be best fit model to forecast ED arrivals in UnityPoint Health, Des Moines. This model can be used by other hospitals with a similar ED arrival pattern as UnityPoint Health to predict hourly ED arrivals. Time series forecasting using ARIMA can be utilized as a decision support system in the healthcare industry. This model can be applied to the ED census and discharge data for deeper analysis of overcrowded EDs.

## Summary points

*7.1 Known facts*

- Monthly and yearly patient arrivals are well established in many research papers.
- Time series forecasting helps emergency departments manage their resources.
- Patient arrival rate varies based on geographic locations and is also influenced by external factors such as weather, transportation, epidemic, and hospital reputation.

*7.2 Our contribution*

- A highly accurate, hourly arrival forecasting model and tests for model fitness was developed.
- Unlike monthly or yearly forecasting models, an hourly forecasting model provides us with a more accurate patient arrival rate and enables the emergency department to schedule their staff.
- Hourly forecasting helps in minimizing ED overcrowding by enabling the ED staff to take proactive actions such as preparing beds and initiating fast track triage for patients with acuity levels of 4 and 5.
- Our study considered both appointed and direct ED arrivals. It also considered ambulatory and arrivals via ambulance.